\documentclass{aa}  

\usepackage{graphicx}
\usepackage{txfonts}
\bibpunct{(}{)}{;}{a}{}{,} 

\begin{document}

\title{\texttt{ellc} - a fast, flexible light curve model for detached eclipsing binary
stars and transiting exoplanets.}

\author{P. F. L. Maxted}
\institute{Astrophysics Group, Keele University, Staffordshire ST5 5BG, UK\\
\email{p.maxted@keele.ac.uk}
}

\date{Received --; accepted --}

 
\abstract
{Very high quality light curves are now available for thousands of detached
eclipsing binary stars and transiting exoplanet systems as a result of surveys
for transiting exoplanets  and other large-scale photometric surveys.}
{I have developed a binary star model (\texttt{ellc}) that can be used to
analyse the light curves of detached eclipsing binary stars and transiting
exoplanet systems that is fast and accurate, and that can include the effects
of star spots, Doppler boosting and light-travel time within binaries with
eccentric orbits. }
{The model represents the stars as triaxial ellipsoids. The apparent flux from
the binary is calculated using Gauss-Legendre integration over the ellipses
that are the projection of these ellipsoids on the sky. The model can also be
used to calculate the flux-weighted radial velocity of the stars during an
eclipse (Rossiter-McLaghlin effect). The main features of the model have been
tested by comparison to observed data and other light curve models. }
{ The model is found to be accurate enough to analyse the very high quality
photometry that is now available from space-spaced instruments, flexible
enough to model a wide range of eclipsing binary stars and extrasolar
planetary systems, and fast enough to enable the use of modern Monte Carlo
methods for data analysis and model testing.}
{}


\maketitle
%

\section{Introduction}
 The discovery of transiting extrasolar planets at the start of this century
has motivated considerable efforts to produce very high quality light curves
that can be used to discover and study these systems. Instrumentation,
observing techniques and data analysis methods for ground-based observations
have all been improved so that its is now possible to detect eclipses with
depths as small as 600ppm  on individual targets
\citep{2015arXiv150602471D}, while ground-based surveys such as WASP
\citep{2006PASP..118.1407P} and HATNet \citep{2004PASP..116..266B} now
routinely discover transiting extrasolar planets with eclipse depths $\approx
1\%$ from surveys that monitor millions of stars using dedicated robotic
instruments. The CoRoT satellite produced light curves for thousands of stars
with a photometric precision $\la 0.1\%$ during its 6-year mission lifetime
\citep{2009A&A...506..411A, 2013Icar..226.1625M}.  The Kepler mission has
discovered hundreds of transiting extrasolar planets from a survey of
approximately 150,000 stars over its 4-year mission lifetime
\citep{2015ApJS..217...31M}. The quality of the photometry produced by the
Kepler instrument \citep[$\sim 20$ppm precision on a time scale of
6-hours,][]{2010ApJ...713L.120J} is orders-of-magnitude better than that
available prior to this remarkably successful mission. The volume and quality
of data are both set to improve as a result of current and future transiting
planet surveys such as the K2 mission \citep{2014PASP..126..398H}, PLATO
\citep{2014ExA....38..249R} and TESS \citep{2015JATIS...1a4003R}, and other
large surveys such Gaia \citep{2005ESASP.576..569D}.

 These advances have motivated researchers to develop analysis techniques and
models that can exploit the full potential of these very high quality data to
study stars and planets in a way not possible before the advent of photometry
from space. For example, \citet{2002ApJ...580L.171M} presented exact analytic
formulae for the eclipse of a spherical star described by quadratic or
nonlinear limb darkening by a planet or other dark body. This enabled the
development of the Monte Carlo techniques that are now the standard tools for
the  analysis of exoplanet observations \citep{2006MNRAS.373..799C,
2013PASP..125...83E}. 

 The data produced by surveys for transiting extrasolar planets are a bonanza
for studies of all types of variable stars, including eclipsing binary stars.
The Kepler archive alone contains high-quality photometry for 2878 eclipsing
binaries \citep{2015arXiv151208830K}. Microlensing surveys have also been a
fruitful source of new DEBS, e.g., the OGLE-III survey discovered over 11\,000
new eclipsing binary stars from a photometric survey of the Galactic disc
\citep{2013AcA....63..115P} and thousands more have been identified in the
Magallenic Clouds by this survey and other microlensing projects
\citep{2013AcA....63..323P, 2014MNRAS.443..432M}. The analysis of the light
curves for detached  eclipsing binary stars (DEBS) combined with radial
velocity measurements for both stars in the binary make it possible to measure
precise, model-indendent masses and radii for a wide variety of stars, from
white dwarfs in binaries with orbital periods of a few hours
\citep{2014MNRAS.438.3399B} to red giants binaries with orbital periods  of
months or years \citep{2013Natur.495...76P}. These fundamental data for stars
can be used to calibrate empirical mass-radius-luminosity relations and to
test stellar models for normal stars \citep{2010A&ARv..18...67T}, to study the
influence of factors such as rotation, magnetic activity and composition on
the structure of stars \citep{2014ApJ...789...53F}, and to improve age and
distance estimates for stellar clusters in which DEBS reside
\citep{2012A&A...543A.106B}.

 I have developed  a new binary star model that is designed for the analysis
of the light curves  and radial velocity curves of detached eclipsing binary
stars and transiting extrasolar planets. My motivation for developing this
software is to have a tool that is accurate enough to analyse the very high
quality photometry that is now available from space-spaced instruments,
flexible enough to model a wide range of eclipsing binary stars and extrasolar
planetary systems, and fast enough to enable the use of modern Monte Carlo
methods to explore the potentially large parameter spaces that result when
dealing with ``real world'' data sets that include astrophysical,
environmental and instrumental noise sources. The light curve model is called
\texttt{ellc}, and is implemented as fortran subroutines called directly from
a user interface written in python. Here I provide a complete description of
the \texttt{ellc} light curve model and give examples of its application to a
variety of binary star and exoplanet systems. The source code and examples are
available as an open-source software project.

\section{The light curve model}

 For convenience I refer to the two bodies in the binary system as stars, but
the description below applies equally to brown dwarfs or exoplanets.

\subsection{Coordinate systems}

 The shape of each star is defined on a Cartesian coordinate system $(x, y, z)$
where the origin is at the star's centre-of-mass, the $x$ axis points towards
the centre of the companion star, the $y$ axis is perpendicular to the $x$-axis
in the orbital plane and the $z$ axis is parallel to the orbital angular
momentum vector, {\bf L}$_{\rm orb}$. This is a right-hand coordinate system so
the direction of orbital motion is towards the negative $y$ direction.

\begin{figure}
\resizebox{\hsize}{!}{\includegraphics{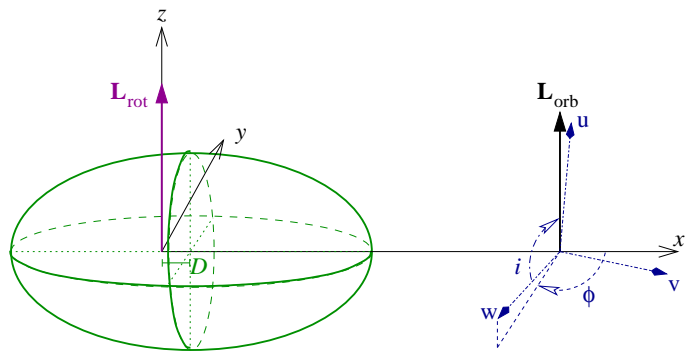}}
\caption{Coordinate systems used to define the shape and position of each
star. The companion is located on the positive $x$-axis and the angular
momentum vector for the orbital motion is parallel to the $z$-axis. The
inclination of the orbital axis to the line of sight is $i$. The origin of the
$(u,v,w)$ coordinate system is at the centre-of-mass of the binary system.
Other symbols are defined in the text.
\label{xyzFig}}
\end{figure}

 The projection of the binary system onto the plane of the sky is described
using a Cartesian coordinate system $(u, v, w)$ with its origin at the binary
centre-of-mass, the $w$ axis pointing towards the observer, and the $u$-axis
parallel to the projection of {\bf L}$_{\rm orb}$ onto the plane of the sky.
The inclination of the binary orbit, $i$, is measured in the normal way, i.e.
the angle of the vector {\bf L}$_{\rm orb}$ to the line of sight.  These
coordinate systems are illustrated in Fig.~\ref{xyzFig}.

 The shape of the star is approximated using a triaxial ellipsoid. The
projection of this triaxial ellipsoid onto the plane of the sky is an ellipse.
The specific intensity distribution over the visible surface of the star is
calculated using a Cartesian coordinate system $(s,t)$ defined by the major and
minor axes of this ellipse, respectively.

All lengths are measured relative to the semi-major axis of the binary star
orbit, $a$. 

\subsection{Star shapes}
 The shape of the star is approximated by a triaxial ellipsoid centred at a position $(x,y,z) = (D,0,0)$
 with semi-axes $(A, B, C)$ aligned with the $x$, $y$ and $z$ axes,
respectively, i.e. \[\frac{(x-D)^2}{A^2} + \frac{y^2}{B^2} +
\frac{z^2}{C^2} = 1.\] All lengths are relative to the semi-major axis of
the binary orbit. This is the approximation that was also used in the WINK
light curve model \citep{1971AJ.....76..701W}. Three methods can be used to
determine the values of $A, B, C$ and  $D$. The simplest model is a spherical
star of radius $R$ centred at the centre-of-mass, for which $A=B=C=R$ and
$D=0$. This model can be very useful for rapidly calculating a large number of
light curves and is accurate enough for many cases. The Roche potential is
widely used to approximate stellar shapes in light curve models. I have used
the definition of this potential described by \citet{1979ApJ...234.1054W},
which includes a modification of the potential to allow for non-synchronous
rotation of the star.  The direction of the star's rotational angular momentum
vector, {\bf L}$_{\rm rot}$ is assumed to be parallel to {\bf L}$_{\rm orb}$.
The values of $A,B,C$ and $D$ are set by requiring that  the value of this
potential is equal at the six points where the ellipsoid intersects the $x$,
$y$ and $z$ axes, i.e. the ellipsoid matches the location of an equipotential
surface at these points. The Roche potential  makes the assumption that the
entire mass of the star is located at the centre of mass. More realistically,
the mass distribution within a star can be calculated by assuming a polytropic
equation of state. The equilibrium shape of a polytrope in the tidal field of
a companion star has been calculated by \citet{1933MNRAS..93..449C}.
Chandrasekhar also calculated the effect of rotation on a polytropic star
\citep{1933MNRAS..93..390C} and discussed how, to a good approximation,
rotation and tidal distortion can be treated independently to describe the
shape of a star in a binary system \citep{1933MNRAS..93..462C}. I take the
same approach and use Chandrasekhar's calculations for the tidal distortion of
a polytrope. For the rotational distortion the model interpolates the results
tabulated by \citet{1964ApJ...140..552J} for polytropes with polytropic index
$n=1.5$ or $n=3$. A polytrope with $n=1.5$ is a  good approximation for stars
where energy transport in the outer layers is dominated by convection and for
gaseous planets, $n=3$ is more appropriate for stars whose structure is
dominated by radiation pressure. The offset of the star towards the companion
is $D = \Delta_3 q \left(R/d\right)^4\left(R/a\right)$ where the coefficient
$\Delta_3$ is taken from Table VI of \citet{1933MNRAS..93..449C}, $d$ is the
distance between the stars' centres-of-mass, and $q$ is the mass ratio. The
coefficients $\xi_p$ and $\xi_e$ from Table 1 of \citet{1964ApJ...140..552J}
are used to calculate the polar and equatorial radii of an oblate spheroid
centred at $(x,y,z) = (D,0,0)$ with its minor axis parallel to {\bf L}$_{\rm
rot}$. I then apply the model for the tidal distortion of this surface given
in equation (35) of \citet{1933MNRAS..93..449C} to this oblate spheroid, again
assuming that {\bf L}$_{\rm rot}$ is parallel to {\bf L}$_{\rm orb}$. 

 For eccentric orbits I model the star using the equilibrium shape at each
point in the orbit, i.e. I do not include any dynamical tidal effects in the
model. The shape of the star will vary through the orbit as a result of the
variations in the tidal field due to the companion but the volume is assumed
to be  constant \citep{1933MNRAS..93..449C}.

\subsection{Star positions}
 
 Times are measured relative to a reference time $t_0$ when the separation of
the stars' centres-of-mass projected on the sky is at a minimum. The two stars
in the binary are labelled 1 and 2 with star 1 furthest from the observer at
time $t_0$, i.e. star 1 will be eclipsed by star 2 at time $t_0$ if the
inclination of the binary is $i=90^{\circ}$. If star 1 is hotter than star 2
and the orbit is approximately circular then this eclipse will be the primary
eclipse, i.e. deeper than the secondary eclipse that will occur approximately
half an orbit later. I assume that the stars follow Keplerian orbits with fixed
orbital eccentricity, $e$. At time $t_0$ the  longitude of periastron for star
1 is $\omega_0$ and the orbital inclination $i_0$. Apsidal motion can be
included by specifying a non-zero value for $\dot{\omega} =
\frac{d\omega}{dt}$, so that the longitude of periastron for star 1 at time
$t_i$
is $\omega_{\rm 1} = \omega_0 + \dot{\omega}t_i$ and the longitude of periastron
for star 2 is $\omega_{\rm 2} = \omega_{\rm 1}+\pi$. Similarly, the inclination
of the orbit at time $t_i$ is assumed to be $i = i_0 + \frac{di}{dt}t_i$. The
longitude of periastron  is not defined for circular orbits, in which case I
fix $\omega_0 =0$. 

 At time $t_0$ the true anomaly of star 1 is $\nu_{{\rm 1},0} \approx \pi/2 -
\omega_0$. The exact value of $\nu_{{\rm 1},0}$ is calculated by finding the
value of $\nu_{{\rm 1},0}$ that minimises the projected separation between the
stars' centres-of-mass \citep{1992AJ....104.2213L}.  At other times the
position of star 1 is calculated using Kepler's equation, $M=E-e\sin E$, to
find the eccentric anomaly $E$ from the mean anomaly $M=2\pi(t_i-t_0)/P_a$, where
the anomalistic period $P_a$ is assumed to be constant. The true anomaly of
star 1 is then \[\nu_{\rm 1} =
2\tan^{-1}\left[\sqrt{\frac{1+e}{1-e}}\tan(E/2)\right],\] and  $\nu_{\rm 2} =
\nu_{\rm 1} +\pi$ for star 2. The angle from the $x$-axis towards the
projection of the line of sight in the $x$-$y$ plane for star 1 is then $\phi =
\nu_{\rm 1} - \pi/2 + \omega$ measured clock-wise looking towards the origin
along the $z$-axis (Fig.~1), and similarly for star 2.

 The separation of the centres-of-mass at time $t_i$ relative to the semi-major
axis of the orbit is \mbox{$d = 1 - e\cos(E)$} so the position of star 1
in the observer's $(u,v,w)$ coordinate system defined previously is
$(u_1,v_1,w_1)$, where
\[
\begin{array}{l}
u_1 = d_1\left(\cos\nu_1\cos\omega_1-\sin\nu_1\sin\omega_1\right), \\
v_1 = d_1\cos i\left(\sin\nu_1\cos\omega_1+\cos\nu_1\sin\omega_1\right), \\
w_1 = d_1\sin i\left(\sin\nu\cos\omega_2+\cos\nu_1\sin\omega_2\right), 
\end{array}
\] 
$d_{\rm 1} = dq/(1+q)$, and $q = M_2/M_1$ is the mass ratio. The apparent
position on the sky of star 1 at time $t_i$ accounting for the light travel
time across the orbit is given to a very good approximation by the actual
position of star 1 at time $t_{\rm 1} = t_i + w_1a/c$, where $c$ is the speed
of light and $a$ is the semi-major axis of the binary orbit. The light travel
time across the orbit can be ignored by setting $a=0$, but in this case the
Doppler boosting effects discussed below with not be calculated. The position
of star 2 is calculated in a similar way using $\nu_{\rm 2} = \nu_{\rm 1}+\pi$
and $r_{\rm 2} = d/(1+q)$. If $a>0$ then equation (25) from
\citet{2015MNRAS.448..946B} is used to calculate a correction to the value of
$t_0$ for the light travel time across the orbit so that the apparent time of
mid-eclipse occurs at the time $t_0$ specified by the user.

\subsection{Specific intensity distribution}

  The specific intensity of the point $(s,t)$ on the stellar disc is given by 
\[{\cal I}_{\lambda}(s,t) = {\cal I}_0U(\mu)G(s,t) + H(s,t)U_H(\mu),\]
where ${\cal I}_0$ is a constant for each star, $U(\mu)$ is the limb darkening
law, $G(s,t)$ accounts for gravity darkening and  the term $H(s,t)U_H(\mu)$
accounts for the irradiation of the star by its companion.

 I have implement several limb-darkening laws to account for the variation of
specific intensity with the cosine of the viewing angle, $\mu$. The simplest
limb-darkening law is linear limb darkening, which depends only on a
single limb darkening parameter, $u$ \citep{Schwarzschild1906}: \[U(\mu) = 1 -
u(1-\mu).\] Two-parameter limb-darkening laws implemented in \texttt{ellc}
include a square-root limb-darkening law \citep{1992A+A...259..227D} \[ U(\mu)
= 1 - a_1(1-\mu) - a_2(1-\sqrt{\mu}),\] an exponential limb-darkening law
\citep{2003A+A...412..241C} \[ U(\mu) = 1 - a_1(1-\mu) - a_2/(1-e^{\mu}),\]  a
logarithmic limb-darkening law \citep{1970AJ.....75..175K} \[ U(\mu) = 1 -
a_1(1-\mu) - a_2\mu\ln\mu,\] and a quadratic limb-darkening law
\citep{1950HarCi.454....1K}\[ U(\mu) = 1- c_1(1-\mu) - c_2(1-\mu)^2.\]
\citet{2000A+A...363.1081C} introduced the following four-parameter limb
darkening law that has also been implemented: \[ U(\mu) = 1 -
\sum_{j=1}^{4}a_j(1-\mu^{j/2}).\] The 3-parameter limb-darkening law defined
\citep{2009A+A...505..891S} which is equivalent to Claret four-parameter with
$c_1= 0 $ has also been implemented.
 
 For gravity darkening I assume that the specific intensity can be related to
the local gravity by a power law with exponent $y(\lambda)$. Note that
$y(\lambda)$ is a wavelength dependent quantity, not the bolometric gravity
darkening exponent often used in other light curve models. Appropriate values
of $y(\lambda)$ and limb-darkening coefficients for various passbands can be
found in \citet{2011A+A...529A..75C}. The local gravity can be calculated
using the gradient of the Roche potential at any given point on the surface,
but this calculation has a significant impact of the speed of the program if
it is done for every integration point used in the calculation. Instead, 
by default the program calculates the gradient of the Roche potential at the
four points $(x,y,z) = (D+A,0,0)$, $(D-A,0,0)$, $(D,B,0)$ and $(D,0,C)$ and
then uses a simple function to interpolate the value of the surface gravity at
other points on the stellar surface.  An option is provided to use the
point-by-point calculation of the local surface gravity so that the impact of
this approximation can be quantified. For example, in the case of KPD\,1946+4340
shown below we used this option to check that using the default method changes
the light curve by less than 250\,ppm at all phases.

 Irradiation of the star by its companion can be difficult to deal with because
the incident energy of the companion changes the thermal structure of its
atmosphere. As a result, the emergent spectrum can be very different to that
of the incident radiation, e.g., ultraviolet photons from the companion can be
absorbed and re-emitted as emission lines at optical wavelengths. The change
in thermal structure also changes  the local limb-darkening law
\citep{1985A+A...147..281V}. In extreme cases, the incident radiation may
produce an optically thin layer in the upper layer of the star's atmosphere
that will then appear limb-brightened.  If the incident radiation is absorbed
and re-emitted with a different spectrum then a complete physical model would
require as input an accurate estimate of the incident flux and its spectrum.
This spectrum may be difficult to observe and hard to predict accurately from
models, particularly at ultraviolet wavelengths where the majority of the flux
from hot stars is emitted, interstellar absorption is severe and the models
are strongly affected by line-blanketing effects. In principle, calculating
the spectrum of the incident radiation at each point on the star's surface
would require integration of the emergent spectrum over the visible surface of
the companion accounting for the variation in limb darkening and gravity
darkening with wavelength, and then iterating to account for the radiation
re-emitted by the star back to its companion. For exoplanets the situation
becomes even more complicated because of the possibilty of advection, i.e.
the incident energy may not be emitted at the location where it is absorbed.
As a result, implementing a prescription for irradiation based on physical
models in a general purpose light curve model would lead to a large
computational overhead,  the results may then depend on model parameters in a
way that is difficult for the user to understand and control, and the results
may be no more accurate than a more simplistic approach. For these reasons I
have implemented a simple prescription for irradiation based on three
parameters that relate directly to specific intensity distribution on the
surface of the star and its angular dependence. In cases where an accurate
physical model for irradiation is available, e.g., simple reflection by
scattering from free electrons, it is then possible to set the values of these
parameters appropriately. For cases of weak irradiation, the simplicity of the
model will not be an issue because the effect on the resulting light curves
will be small. If the observed light curves are clearly  affected by
irradiation then some or all of the parameters can be included as free
parameters in the fit in order to explore how much the parameters of interest
are affected by the way irradiation is dealt with, or to gain some insight
into the physics of irradiation.

 The specific intensity of the point $(s,t)$ includes the term
$H(s,t)U_H(\mu)$ to deal with irradiation. The factor $U_H(\mu)$ is a linear
limb darkening law with coefficient $u_H$. The angle between the local surface
normal and the vector from the point $(s,t)$ to the centre of the companion,
$\gamma$, is calculated by assuming that the surface of the ellipsoid can be
approximated by a sphere with the same radius as the distance of the point
from the centre of the ellipsoid, $r(s,t)$.  The distance from the point
$(s,t)$ to the centre of the companion is $d$ and the companion is assumed to
be a sphere of radius $r_c = \left(ABC\right)^{\frac{1}{3}}$. If $\cos\gamma >
\Delta = \left(r(s,t)-r_c\right)/d$ then the entire disc of the companion is
visible so I take the irradiating flux to be \[{\cal F}_{\rm irr}(s,t) =
\left\{ \begin{array}{lr} {\cal F}_{\rm 0,c}/d^2 & \cos\gamma > \Delta, \\
{\cal F}_{\rm 0,c} (\cos\gamma+\Delta)/ \left(2|\Delta|d^2\right) &
|\cos\gamma| < \Delta, \\ 0  & \cos\gamma < -\Delta, \end{array} \right. \]
where ${\cal F}_{\rm 0,c} $ is the specific intensity integrated over the
visible hemisphere of the companion for a distant observer viewing the
companion along the $x$ axis. This integration is done using equation
(\ref{integral}) discussed below. The specific intensity normal to the surface
is taken to be \[H(s,t) = H_0{\cal F}_{\rm irr}(s,t)^{H_1} \] Several
geometrical factors of order 1, albedo effects and the redistribution of
incident energy into the observed wavelength region are all subsumed into the
parameter $H_0$. The parameter $H_1$  can be used to give some control over
how the specific intensity depends on the strength of the irradiation. Using
$H_1\gg 1$ will create a ``hot spot'' near the point closest to the companion,
whereas $H_1=0$ will result in approximately uniform brightness for all points
with $\cos\gamma > \Delta$.

\subsection{Integration}
 To obtain the observed flux the specific intensity must be integrated
over the visible area of the star. This integration is implemented using a
combination of numerical Gaussian-Legendre quadrature and exact analytical
expressions for the areas of overlapping ellipses. The integral
required can be written 
\begin{equation}
F_{\lambda} = \int\!\!\!\int {\cal I}_{\lambda}(s,t) df dg \approx 
\sum_{j=1}^{N_{j}} w_j {\cal I}_{\lambda}(s_j,t_j) 
\frac{\int\!\!\!\int df dg }{\sum_{j=1}^{N_{j}} w_j}. 
\label{integral}
\end{equation}
The first summation in this expression represents Gaussian-Legendre quadrature
over a two-dimensional non-rectangular grid of $N_j$ points at locations
$(s_j,t_j)$  with weights $w_j$. The size of the grid is specified by the
number of grid points along the major axis on the ellipse. In the current
implementation this value is either 4, 8, 16, 24 or 32. The distribution of the
integration points is approximately uniform over the visible area. The
summation in the denominator of the ratio is an estimate of the area of the
visible stellar disc calculated in the same way. The actual area of the
stellar disc that is visible is given by the integral in the numerator of the
ratio, and can be calculated exactly since it is either the area of an
ellipse, or the difference between this area and the area common to two
overlapping ellipses.  The ratio of the integral and sum has a fixed value for
a given combination of  $N_j$, $(s_j,t_j)$ and $w_j$, and serves as a
correction that improves the accuracy of the numerical integration. An
alternative view of the same expression is that flux is calculated from the
visible area of the ellipse, which can be calculated exactly, weighted by the
average intensity over the visible area, which is calculated by numerical
integration. The calculation of the overlap area of two ellipses requires a
very robust algorithm to calculate the number and positions of the
intersections between two ellipses. The algorithm I have developed is
described in Appendix A.

\subsection{Flux scale and surface brightness ratio}
 The surface brightness ratio is an ambiguous quantity for limb-darkened stars
in a binary that  do not emit isotropically. To define a flux scale in
\texttt{ellc} I first set the specific intensity normal to the surface at the
point on star 1 closest to its companion to 1, i.e. ${\cal I}_{\rm 0,1}=1$. I
then use equation (\ref{integral}) to calculate ${\cal F}_{\rm 0,1}$, the
integral of the surface-brightness distribution for the star 1 for a distant
observer viewing the star along the $x$-axis. This calculation step ignores
eclipses, spots and heating effects by the companion, but does include the tidal
distortion terms. The same calculation for star 2 with ${\cal I}_{\rm 0,2}=1$
gives the quantity ${\cal F}_{\rm 0,2}({\cal I}_{\rm 0,2}=1)$ which is then
used to set the value of ${\cal I}_{\rm 0,2}$ used in all subsequent steps in
the calculation  such that \[ {\cal I}_{\rm 0,2} = S_{\lambda} {\cal F}_{\rm
0,1}/{\cal F}_{\rm 0,2}({\cal I}_{\rm 0,2}=1),\] where $S_{\lambda}$ is the
value specified by the user for the disc-averaged surface brightness ratio at
the wavelength of observation $\lambda$.

 The next step in the calculation is to set a normalising factor for the
output light curves. This is done in the same way as the calculation for
${\cal F}_{\rm 0,1}$ but in this case the irradiation factor is included. This
calculation yields two normalising factor ${\cal F}_{\rm N,1}$ and ${\cal
F}_{\rm N,2}$. The calculation of the light curve then proceeds using equation
(\ref{integral}) to calculate the ${\cal F}_{i,1}$ and ${\cal F}_{\rm
i,2}$, the apparent fluxes at times $t_i$ for stars 1 and 2, respectively,
including irradiation, eclipses, star spots and Doppler boosting.  The values
of ${\cal F}_{i,1}$ and ${\cal F}_{i,2}$ can be obtained directly
using the routine \texttt{fluxes} in the pyhton module \texttt{ellc}. The
light curve generated by the routine \texttt{lc} in the same module returns
the value
\[{\cal F}_{i} = \frac{{\cal F}_{i,1} + {\cal F}_{i,2} + {\cal F}_3}{
  {\cal F}_{\rm N,1} + {\cal F}_{\rm N,2}+ {\cal F}_3},\] where
${\cal F}_3 = \ell_3\left({\cal F}_{i,1} + {\cal F}_{i,2}\right)$ and
$\ell_3$ is a value for ``third-light'' specified by the user.

\subsection{Simplified reflection effect}
 Integrating the illumination pattern on one star by the other  can result in
severe numerical noise in the light curve due to the sharp boundary between
the illuminated and non-illuminated hemispheres if a sparse integration grid,
must be used, e.g., if a large number of approximate light curves need to be
calculated to explore the parameter space of a least-squares fitting problem.
To avoid this problem a simplified model for the reflection effect is 
available which is useful in cases where the amplitude of the reflection
effect is low. In this simplified model the flux of each star increased by an
amount 
\[ {\cal R}_{1,2} = \alpha_{1,2}{\cal I}_{0,(2,1)}(R_{1,2}/d)^2\left[
  \frac{1}{2} + \frac{1}{2}\sin^2i\sin^2\phi_{1,2} \pm \sin
i\sin\phi_{1,2}\right],\]
where the positive and negative cases of the final term apply to stars 1 and 2,
respectively. During the eclipse this quantity is reduced by a factor equal to
the fractional loss of flux from the eclipse star. This is the same model for
the reflection effect used in {\sc ebop}  \citep{1981psbs.conf..111E} except
that {\sc ebop} includes a factor $(R_{1,2}/a)^2$  in ``geometric reflection
coefficient'' whereas I use the factor $(R_{1,2}/d)^2$ explicitly in the
model.

\subsection{Radial velocity}
 If the semi-major axis of the binary orbit ($a$) is set to a non-zero value
then the radial velocity ($v_r$) of each star is calculated. The
centre-of-mass radial velocity is calculated from the Keplerian orbit of the
star and can be obtained using the routine \texttt{rv} in the \texttt{ellc}
python module. A check is made in the program for a non-zero value of $a$ and
an orbital period value $P_a=1$, which is assumed to be an error in input
because $P_a=1$ is used to denote that the input times for calculation are in
phase units, not days as required for the correct calculation of the radial
velocity. 
 
 The default behaviour in the current version of \texttt{rv} is to return
radial velocity values that are weighted by the flux from every point on the
visible surface of the star (``flux-weighted radial velocity''). These are
calculated in the same way as the flux values used for the calculation of the
light curves. The projected rotational velocity at every point on the star's
surface can be calculated using the asynchronous rotation factor, $F_{\rm rot}
= P_a/P_{\rm rot}$, that is also used to define the shape of the star with
rotation period $P_{\rm rot}$. Alternatively, the projected equatorial
rotation velocity at the stars' equator, $V_{\rm rot}\sin i$, can be used to
determine the projected rotational velocity over the stellar surface. The
effect of star spots on the flux-weighted radial velocities is acccounted for
by multiplynig the flux-deficit due to the spot by the  projected rotational
velocity at its centre. This approximation will not be accurate for large
spots.

 The effect of eclipses on the flux-weighted radial velocity is accounted for
and so \texttt{ellc} can be used to model the Rossiter-McLaughlin (R-M) effect
for stars in which the rotation axis is aligned with the orbital axis. It is
also possible with \texttt{ellc} to model the R-M effect for a spherical star
eclipsed by star or planet described by the Roche potential, a polytrope or a
sphere for which the projection of the rotation axis on the sky during the
eclipse is at an angle $\lambda$ to the orbital axis, as defined by
\citet{2005ApJ...622.1118O}.

\subsection{Doppler boosting}
 The flux-weighted radial velocity can be used to correct the apparent flux for
the effects of Doppler boosting, which increases the flux by a factor
$(1-v_r/c)^3$, and the Doppler shift, which reduces the flux by a factor
$\approx (1-v_r/c)^2$ that depends on the wavelength gradient of the stellar
spectrum \citep{2000MNRAS.317L..41M}. For radial velocities $v_r\ll c$ the two
effects can be combined into a single factor $1-B\frac{v_r}{c}$, where $B = 5+
\frac{d\ln F_{\lambda}}{d \ln \lambda}$ is calculated separately
\citep{2011MNRAS.410.1787B}, or can be set to zero to ignore these effects.  
The flux-weighted radial velocity accounts for the eclipse of the star and so
\texttt{ellc} can be used to model the photometric R-M effect (see section
\ref{PhotRMSection}, below). Using centre-of-mass radial velocities with
Doppler boosting gives inconsistent results for the eclipses in the light
curves  (because the photometric R-M effect is not included) and so is
not recommended.

\subsection{Star spots}
 The effect of circular spots on the light curve of a spherical star with
quadratic limb-darkening have been calculated by \citet{1994ApJ...420..373E}
and the resulting integrals are provided in convenient form in the appendix to
that paper together with an erratum \citep{1994ApJ...430..438E}.  I use these
equations in \texttt{ellc}  to calculate the variation in flux from each star
due to spots in the model. In cases where the limb-darkening law used is not
quadratic I set the limb-darkening coefficients used in the spot
model so that it  matches the actual limb-darkening law
used at the points $\mu = 0, 0.5$ and 1. In cases where the spot  is eclipsed
by the companion I reduce the effect of the spot by a factor equal to the
fraction of the spot covered by the companion.  The position of the centre of
the spot on the sky is calculated based on its longitude and latitude on the
triaxial ellipsoid and its shape is calculated by approximating the surface of
the triaxial ellipsoid by a sphere with the same radius as the distance from
the centre of the spot to the centre of the ellipsoid. The calculation of the
eclipsed area is straightforward if it is not on the limb of the star because
the projected shapes of the spot and the companion are both ellipses. The
calculation is more difficult if the spot intersects the limb of the star
because it involves the intersection of two ellipses and a circle. The area
required can be found as the sum of various ellipse segments and triangles so
is quick to calculate, but this does require some effort to implement as there
are approximately 90 different possible configurations for the two ellipses
and the circle that must be considered.

\section{Examples}

 In this section I test the reliability of \texttt{ellc} by comparing it to
other binary star models and show how \texttt{ellc} can be used to analyse
light curves and radial velocity observations for eclipsing binary star
systems using Monte Carlo methods.

\begin{figure}
\resizebox{\hsize}{!}{\includegraphics{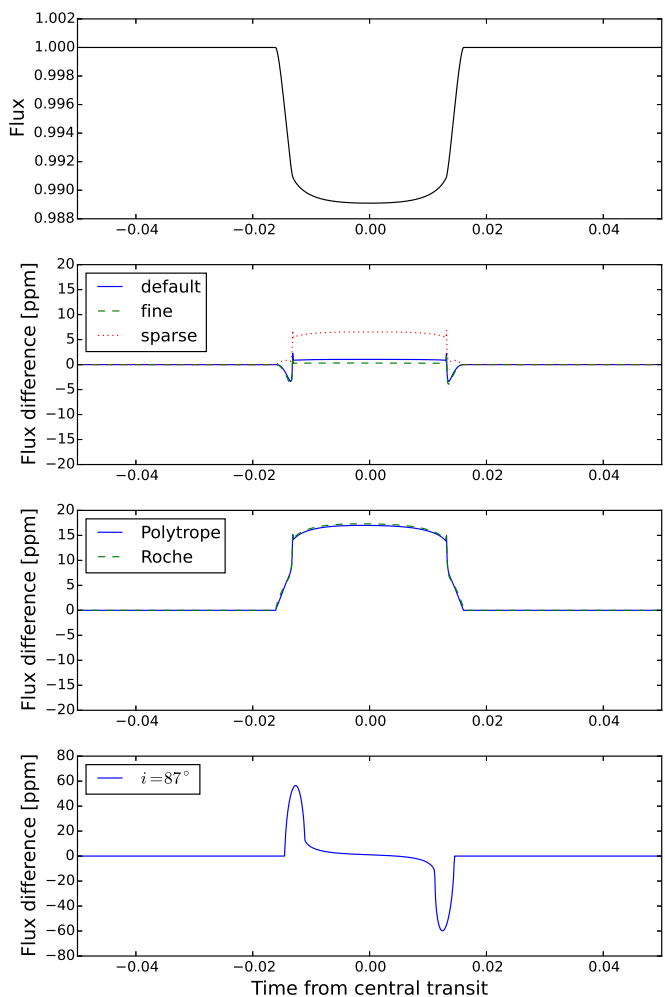}} \caption{{\bf Upper panel}
Light curve for a planet on an eccentric orbit with $i=90^{\circ}$ calculated
using \texttt{batman}. {\bf Upper-middle panel} Difference between the light
curves calculated using  \texttt{ellc} for a spherical planet with the grid
sizes indicated and with \texttt{batman} for the case $i=90^{\circ}$. {\bf
Lower-middle panel} Difference between the light curves calculated using
\texttt{ellc} with the default grid size for planet described by either a
Roche potential or a polytrope with index $n=1.5$ and with \texttt{batman} for
the case $i=90^{\circ}$. {\bf Lower panel} Difference between the light curves
calculated using  \texttt{ellc} with the default grid size and a spherical
planet and with \texttt{batman} for an eccentric orbit with $i=87^{\circ}$.
The difference in flux of approximately 50ppm during ingress and egress is the
result of an inaccurate estimate for the time of periastron in the
\texttt{batman} model. \label{batman}}
\end{figure}

\subsection{\texttt{batman}}
\citet{2015PASP..127.1161K} has developed a python package for modeling
exoplanet transit and eclipse light curves with an algorithm that provides
very high numerical precision. I have used \texttt{batman} to evaluate the
numerical precision of the light curves calculated using \texttt{ellc}. The
upper panel of Fig.~\ref{batman} shows the light curve calculated using
\texttt{batman} version 2.1.0 for the case of the transit of a spherical star
by a spherical planet with a radius of 0.1 stellar radii ($R_p =
0.1\,R_{\star}$). The semi-major axis of the orbit is 10\,$R_{\star}$ and
the inclination of the orbit is $i=90^{\circ}$. The quadratic limb-darkening
coefficients are $c_1=0.1$ and $c_2=0.3$. The eccentricity of the orbit is
$e=0.1$  and the longitude of periastron is $\omega = 60^{\circ}$. 
The next panel down in this figure shows the difference between this light
curve and light curves calculated using \texttt{ellc} with a grid size of 8
(``sparse''), 16 (``default'') or 24 (``fine''). The models agree to better
than 10ppm at all phases of the transit for the sparse grid and to within a
few ppm for the default and fine grids.  This level of precision is more than
sufficient given the level of systematic error due to astrophysical phenomena
not included in these models. To give just one example, the next panel down in 
Fig.~\ref{batman} shows the difference between the same light curve calculated
with \texttt{batman} and light curves calculated using \texttt{ellc} using
either a Roche potential or a polytrope with $n=1.5$ to calculate the shape
of the planet assuming that the mass ratio is $q=0.001$. The difference of
about 15ppm seen in this panel is a due to the systematic error in light curve
calculated with \texttt{batman}  that arises from using a sphere to
approximate the shape of a tidally disorted planet. The current version of
\texttt{batman} does not include the option of calculating light curves for
non-spherical planets. This systematic error can result in the measured radius
being to low by 1\,--\,10\% for a typical hot Jupiter systems
\citep{2011A&A...536C...1L, 2011A&A...528A..41L}.

 I also tested \texttt{ellc} against the results from \texttt{batman} for the
same planetary system viewed at an inclination $i=87^{\circ}$. The results
shown in the lower panel of Fig.~\ref{batman} show differences of about 50ppm
between the two light curves during ingress and egress. Inspection of the
source code for \texttt{batman} shows that the time of periastron relative to
the time of mid-transit is calculated using an expression that assumes
$i=90^{\circ}$. This approximation is not justified at this level of precision
for inclinations $i\ne 90^{\circ}$. This problem was fixed in \texttt{batman} 
version 2.0.0.

\begin{figure*}
\resizebox{\hsize}{!}{\includegraphics{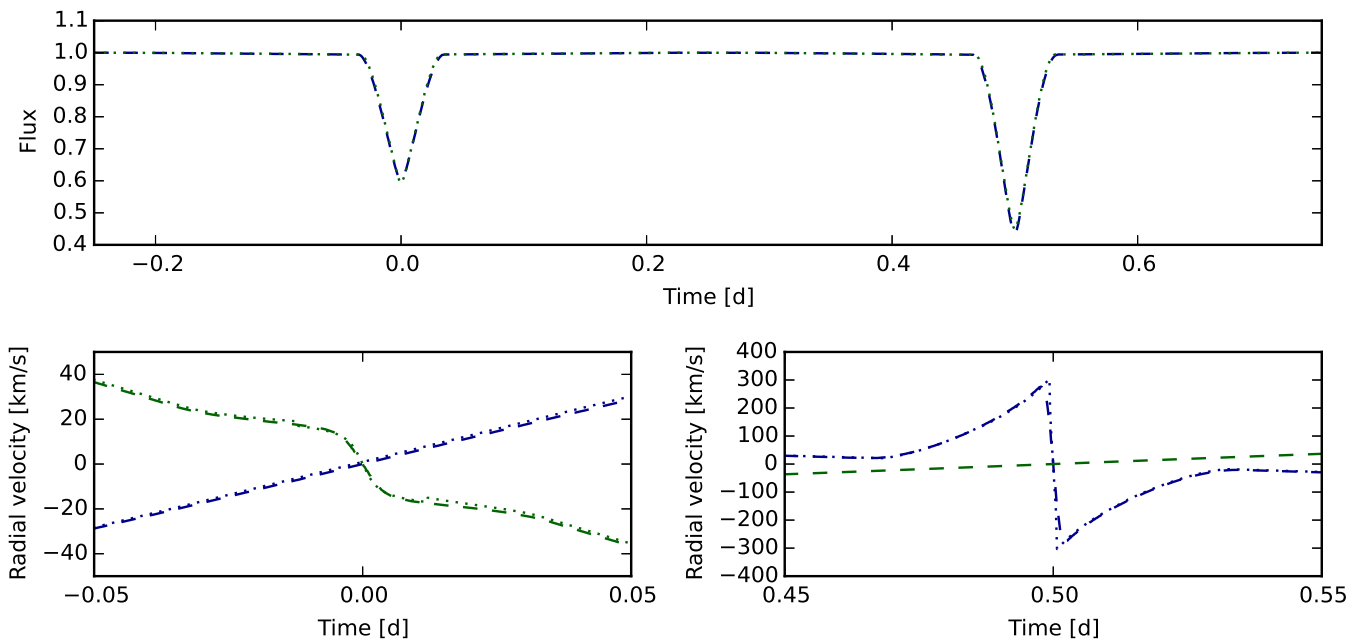}}
\caption{Comparison of the light curves and radial velocity curves for a
binary system showing double-partial eclipses computed using Nightfall (dashed
lines) and \texttt{ellc} (dotted lines). \label{DoublePartial}}
\end{figure*}

\subsection{Double-partial eclipses}
 If two stars in an orbit with $i\approx 90$ are of similar size and one of
the stars rotates close to its break-up velocity, then the rapidly rotating
star will be significantly oblate. This makes it possible for the eclipse of
this oblate star to produce a configuration I call a ``double-partial
eclipse'' in which the poles of the oblate star are eclipsed while the equator
is visible on both sides of the slowly-rotating star. The effect on the light
curve of this configuration is quite subtle, but the Rossiter-McLaghlin (R-M)
effect during the double-partial eclipse has a characteristic large and rapid
change from positive to negative velocity at the mid-point of the eclipse. To
test whether \texttt{ellc} can reproduce this effect correctly I compared
light curves and radial velocity curves for a hypothetical binary system with
double-partial eclipses calculated with \texttt{ellc} to those calculated
using
Nightfall\footnote{\url{http://www.hs.uni-hamburg.de/DE/Ins/Per/Wichmann/Nightfall.html}}
\citep{2011ascl.soft06016W}.  The results are shown in
Fig.~\ref{DoublePartial}. The details of these simulations can be found in the
information provided with the software package, which includes the python
scripts used to generate this plot and the configuration file for Nightfall.
There are small differences between the light curves and radial velocity
curves generated by the two models, but overall the agreement betweed the two
models is very good. The rapid change from positive to negative velocity at
the mid-point of the eclipse is seen very clearly in the radial velocity curve
generated by \texttt{ellc}. 

 The calculation of light curves in 11 bands and radial velocites at 8192
phase points using Nightfall version 1.88  takes 2.3\,s on a MacBook Pro with
a 2~GHz Intel Core i7 CPU. The time taken to calculate a lightcurve in a
single band and 2 radial velocity curves with \texttt{ellc} and to generate
the plot on the same machine is 2.5\,s. I also attempted to use Nightfall to
generate a similar plot for an eccentric orbit but found that there is a bug
in this version of the program that inverts the shape of the R-M effect for
eccentric orbits. The execution time in this case was 113.4\,s. This is much
slower than the calculation  for the circular orbit because Nightfall uses a
grid of points in 3-dimensional space to represent the stars.  There is
a large overhead required to re-calculate the positions of all the points on
this grid at each phase point to account for the varying potential in an
eccentric orbit. There is no significant increase in the execution time for
\texttt{ellc} for an eccentric orbit compared to a circular orbit in this
case.

\begin{figure*}
\resizebox{\hsize}{!}{\includegraphics{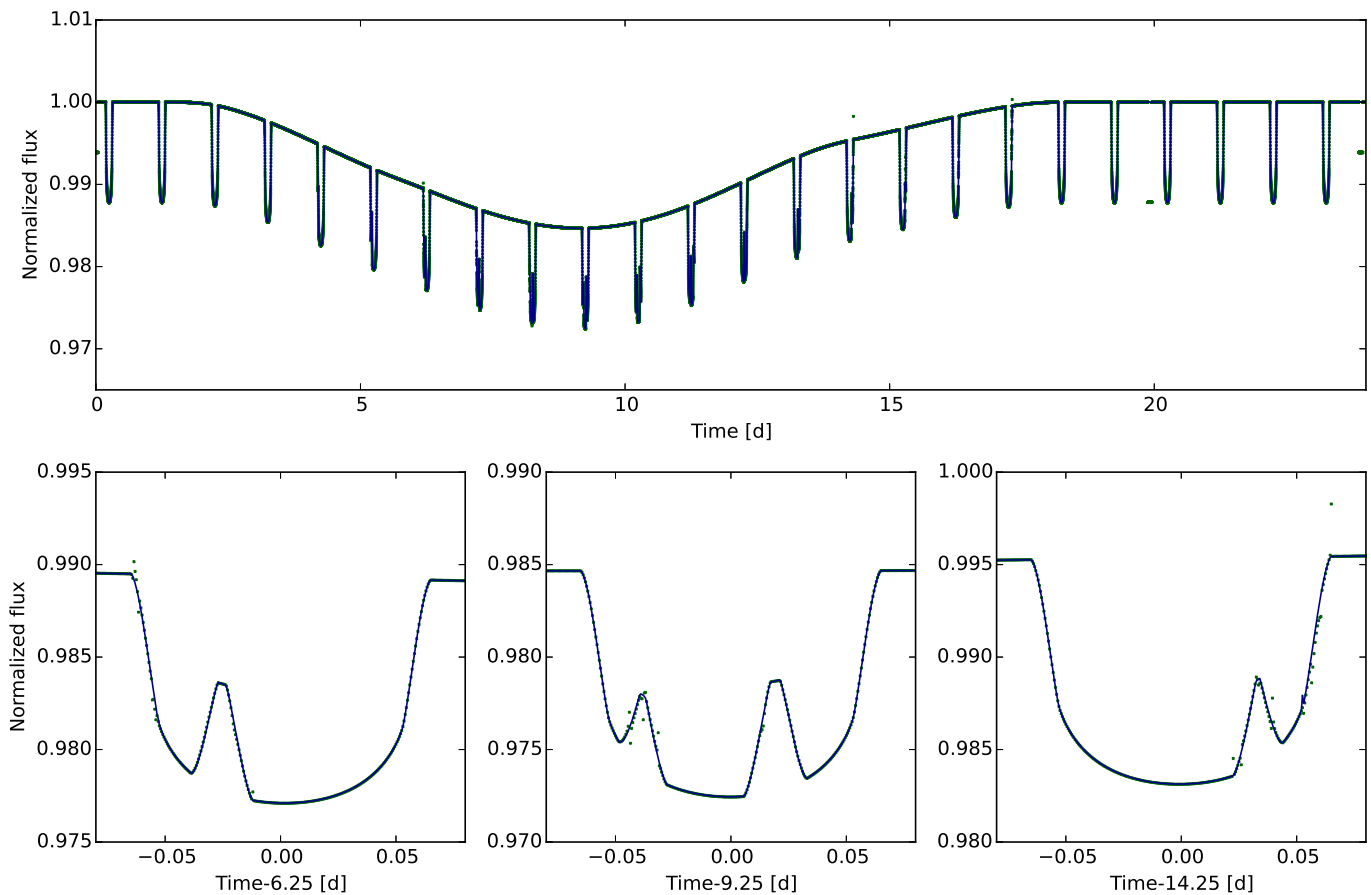}}
\caption{Comparison of the light curves for a spherical star with two circular
spots eclipsed by a spherical dark body calculated using \texttt{ellc} (thin
line) and KSint (points). \label{KSint}}
\end{figure*}

\subsection{KSint}
KSint is a fast numerical algorithm for accurately calculating light
curves for transiting extrasolar planets orbiting spotted stars
\citep{2012MNRAS.420.1630P, 2014MNRAS.444.1721M}. The model assumes that the
star and planet are spherical. As with \texttt{ellc}, the spot profile  is
defined from the interception of a cone with its vertex at the center of the
sphere with the surface of the sphere. I have used \texttt{ellc} and
KSint version 1.0  to calculate the light curve for a spherical star
with a rotation period of 23.9\,days with two dark spots and a planetary
companion on a circular orbit with a period of 1~day and an inclination
$i=90^{\circ}$. The radius of the planet is 0.1 stellar radii ($R_p =
0.1\,R_{\star}$) and the semi-major axis of the orbit is 2.765\,$R_{\star}$.
In reality a planet on such a short-period orbit would be appreciably
non-spherical, but I have assumed that the planet is spherical for the
calculation of the light curve with \texttt{ellc} so that the results are
directly comparable to those calculated with KSint. The limb
darkening coefficients for the quadratic limb darkening law are $c_1 = 0.4$
and $c_2 = 0.3$. The angular radius of one spot ($\approx 5.74^{\circ}$) is
set such that its projected size on the sky is the same as the planet. This
gives a good test of the numerical stability of the algorithm used to
calculate the intersections between the projected edges of the spot and planet.
The radius of the other spot is a factor $\sqrt{2}$ larger. The dimming factor
for both spots is 0.5. Both stars are on the stellar equator and are separated
by an angle of 60$^{\circ}$. The results are shown in Fig.~\ref{KSint}. A few
of the points in the light curve calculated using  KSint are badly
affected by numerical noise, e.g., the low points in the ranges
19.88\,--\,19.95 days and 23.86\,--\,23.94~days, and a few high points near
the first and last contact points of some transits. By contrast, the light
curve calculated using \texttt{ellc} shows very little numerical noise. In
other regions of the light curve, the agreement between the two models is
excellent.

\begin{figure*}
\resizebox{\hsize}{!}{\includegraphics{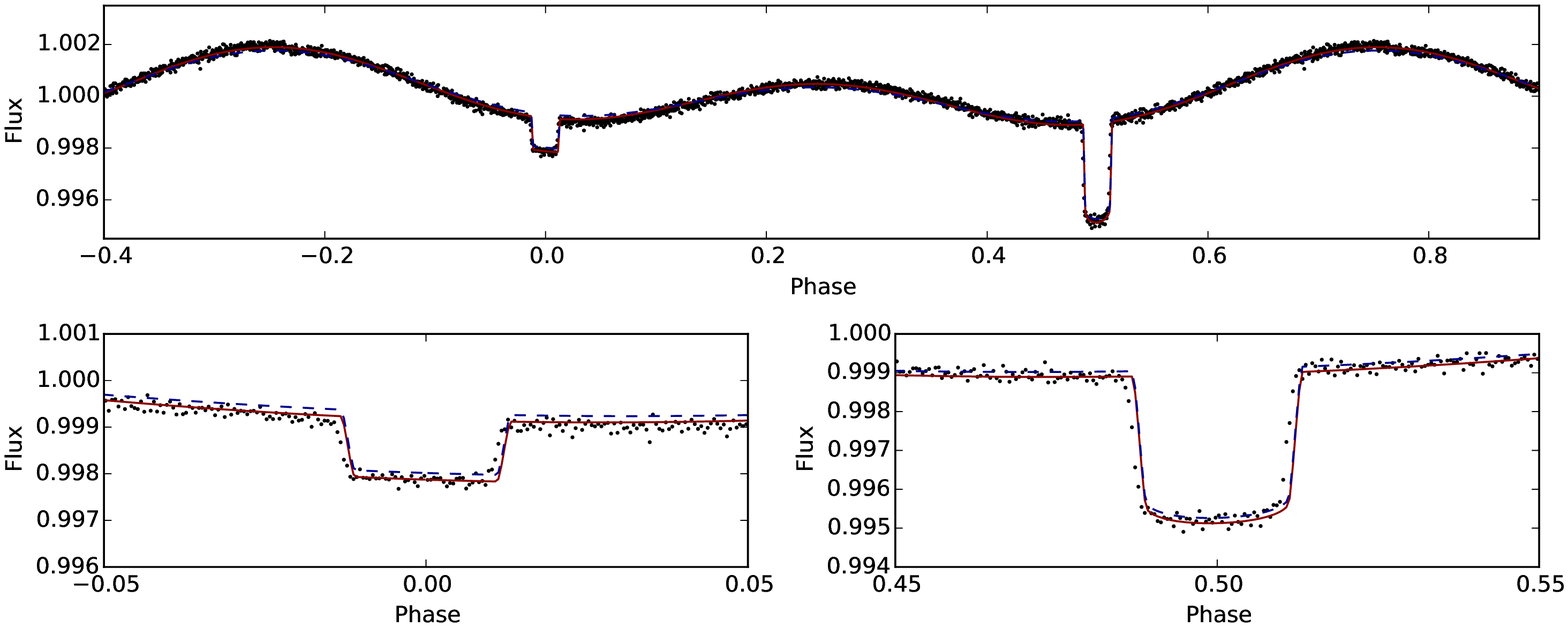}}
\caption{Phase-binned Kepler short-cadence data for KPD\,1946+4340 (points)
compared to light curves generated using \texttt{ellc} using the Roche
potential (solid line) and a polytrope with index $n=1.5$ (dashed line).
 \label{KPD1946+4340}}
\end{figure*}

\subsection{KPD\,1946+4340}
 KPD\,1946+4340 is a subdwarf-B star in a short-period binary system
($P\approx0.4$\,d) with a white dwarf companion \citep{2011MNRAS.410.1787B}.
This configuration together with the very high quality light curve of this
binary obtained by the Kepler mission make KPD\,1946+4340  a useful test case
for the calculation of the ellipsoidal effect and Doppler boosting with
\texttt{ellc}. I downloaded all the observations of KPD\,1946+4340 obtained with
Kepler in short-cadence mode from the Kepler data archive\footnote{\url{
http://archive.stsci.edu/kepler/}} and used the flux values provided in the
column {\sc pdcsap\_flux} of the archive data tables  to create the
phase-binned light curve shown in Fig.~\ref{KPD1946+4340}. This phase-binned
light curve is formed from 1\,007\,160 photometric measurements from Kepler
excluding flagged data points and a few obvious outliers. The ephemeris used
to calculate the phase is from \citet{2011MNRAS.410.1787B} converted to the
Barycentric Kepler Julian date (BKJD) time system used for the Kepler archive
data, i.e. \[{\rm BKJD} =53652.84813(62)+0.40375026(16)E, \]

 The eclipse of the white dwarf occurs at phase 0. The ellipsoidal variation
due to the tidal distortion of the sdB star by the much-fainter white dwarf
companion can be seen as a smooth variation with a period half that of the
orbital period. The Doppler boosting of the light from the sdB star  is also
obvious from the asymmetry in the light curve between phase 0.25 when its is
receeding from the observer and phase 0.75 when it is approaching. These
effects are only obvious because of the superb quality of this light curve --
the rms scatter is approximately 0.01\% and the entire flux range plotted in
Fig.~\ref{KPD1946+4340} is less than 1\%. 

 I used the parameters for this binary system from \citet{2011MNRAS.410.1787B}
to simulate the light curve of KPD\,1946+4340 assuming either a Roche potential
for the stars or a polytrope with index $n=1.5$. Other details of the
simulation can be found by inspection of the python script used to generate
this figure that is included with the software distribution.  The simulated
light curve using the Roche potential gives an excellent match to the light
curve at all phases. In particular, the asymmetry in the light curve due to
Doppler boosting is very well reproduced when the orbital velocity of the sdB
star (which dominates the flux) is set to the value measured from its orbital
Doppler shift. Using a polytrope to describe  the shape sdB star results in a
worse match to the the ellipsoidal variation,  but the parameters used for
this simulation were based on a least-squares fit to the Kepler light curve of
KPD\,1946+4340 using a Roche potenttial, so it may be that a better fit to the
light curve can be achieved for a polytropic model by adjusting the other
parameters in the model. This would certainly be an interesting exercise now
that there is more than 20 times as much  Kepler data available for this
star than was available at the time of the study by
\citeauthor{2011MNRAS.410.1787B}. A full analysis will require very careful
handling of any systematic errors in the data since the signal being analysed
here are very small. This analysis is beyond the scope of this description of
the software, but I demonstrate below that \texttt{ellc} can be used to
analyse high-quality data such as these using Monte Carlo techniques.

\subsection{Photometric R-M effect\label{PhotRMSection}}
\begin{figure}
\resizebox{\hsize}{!}{\includegraphics{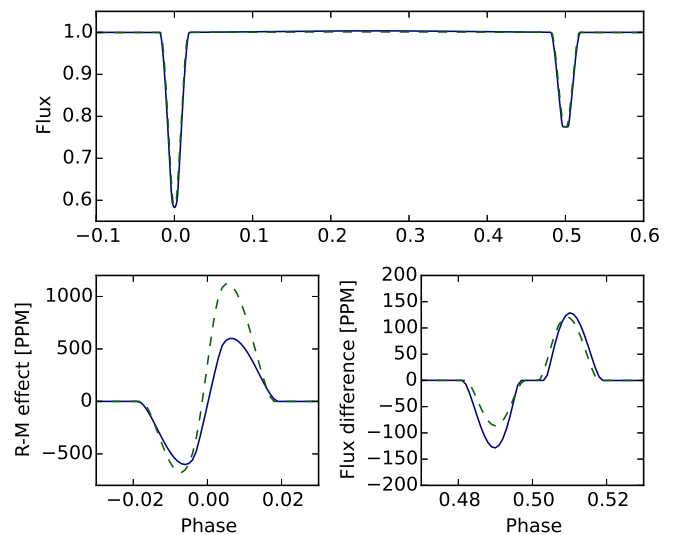}}
\caption{Simulation of the photometric Rossiter-McLaughlin (R-M) effect for a
pair of white dwarf stars rotating close to their break-up velocity in an
eclipsing binary with a period of 39.1 minutes. The upper panels show the
light curve of the binary and the two lower panels show the contribution to
this light curve due to the photometric R-M effect. In all panels, the solid
line is calculated using Roche geometry for stars with rotation axes aligned
with the orbital axis and dashed lines are for models of spherical stars with
$\lambda = 60^{\circ}$.  \label{PhotRM}}
\end{figure}

 The photometric R-M effect is essentially the same phenomenon as the
spectroscopic R-M effect measured using the Doppler shift of stellar
spectral lines. The variations in apparent flux are the result of Doppler
boosting of the light from different parts of a rotating star during an
eclipse \citep{1974A&A....30..135H}. The amplitude of the signal is generally
very small so there are no reported detections of this phenomenon to-date.
\citet{2012ApJ...745...55G} has discussed the prospects for detecting this
signal using simulations that include limb darkening and obliquity but that
neglect the oblateness of the stars. Fig.~\ref{PhotRM} shows a simulation
using \texttt{ellc} for a pair of white dwarfs in a binary system with
parameters similar to those used by \citeauthor{2012ApJ...745...55G}, i.e. an
orbital period of 39.1 minutes, $i=90^{\circ}$ and rotational velocities close
to their maximum possible values. The details of this simulation can be found
by inspection of the python script used to generate this figure that is
included with the software distribution. To estimate the maximum possible size
of the photometric R-M effect \citeauthor{2012ApJ...745...55G} compared the
light curves for the stars rotating at their break-up velocity to stars
rotating synchronously with the orbit. A similar comparison with \texttt{ellc}
shows that the differences in the light curves are completely dominated by the
change in shape of the stars and the change in the surface brightness
distribution due to  gravity darkening. Instead, to generate the flux
difference due to the photometric R-M effect shown in Fig.~\ref{PhotRM} I
calculated the light curves for stars rotating near their break-up velocity
with both prograde and retrograde rotation and have plotted half of the
difference between these results.  The shape and amplitude of the signal are
similar to the results shown by \citeauthor{2012ApJ...745...55G}. Also shown
in Fig.~\ref{PhotRM} is the photometric R-M effect for the same system
assuming that the sky-projected rotation axis is misaligned with the rotation
axis by an angle $\lambda = 60^{\circ}$. This light curve was calculated for
spherical stars because the current version of \texttt{ellc} does not include
misaligned rotation for non-spherical stars.

\subsection{GD448}
\begin{figure}
\resizebox{\hsize}{!}{\includegraphics{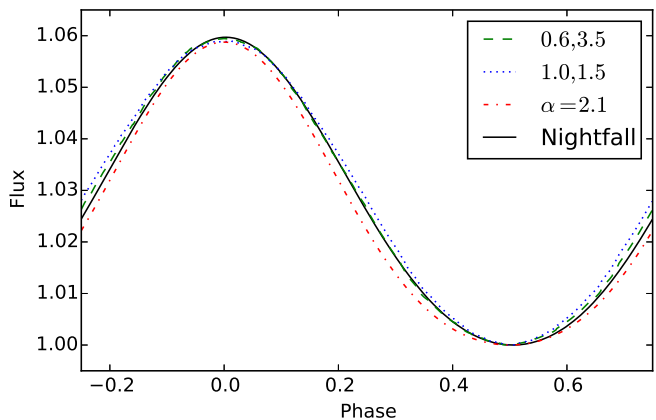}}
\caption{Light curves of the short-period M-dwarf -- white-dwarf binary GD448
simulated using Nightfall and using \texttt{ellc} with two sets of
heating/iraddiation parameters $H_0, H_1$, as indicated, and also using a
simplified model of the reflection effect with $\alpha  = 2.1$.
 \label{GD448}}
\end{figure}

 GD448 is a non-eclipsing,  short-period (P=0.103\,d) M-dwarf -- white-dwarf
binary system \citep{1998MNRAS.300.1225M}. This binary system shows a
reflection effect in the light curve with an amplitude of about 0.1 magnitudes
in the I-band. Fig.~\ref{GD448} shows the I-band light curve of GD448 simulated using
Nightfall using the parameters for the binary from
\citet{1998MNRAS.300.1225M}. The effective temperatures of the stars used in
this simulation were 19\,000\,K and 3\,000\,K, respectively. Further  details
of the simulation can be found by inspection of the python script used to
generate this figure and other files that are included with the software
distribution. The amplitude of the reflection effect is underestimated in
this simulation, partly because the model atmospheres available in Nightfall
are not accurate when applied to a white dwarf. It would be possible to get a
better match the observed amplitude of the reflection effect in GD448
by adjusting the assumed albedo value in the model from its default value of
0.5. Also shown in Fig.~\ref{GD448} are light curves simulated using
\texttt{ellc} with two sets of irradiation parameters $H_0, H_1 = (0.6,3.5)$
and $(1.0,1.5)$. These values were set ``by-eye'' to produce a similar shape
and amplitude to the reflection effect calculated using Nightfall.This
demonstrates that \texttt{ellc} can be used to calculate light curves that
include the reflection effect using reasonable values for the parameters in
the irradiation model but, as with other light curve models,   these free
parameters need to be adjusted to match the shape and amplitude of the
reflection effect in the observed light curves of actual binary stars.
The result of using the simplified reflection effect model is also shown
in Fig.~\ref{GD448} with the reflection coefficient adjusting ``by-eye'' to
match the amplitude of the reflection effect calculated using Nightfall. 

\subsection{1SWASP~J011351.29+314909.7}
 1SWASP~J011351.29+314909.7 (J0113+31) is  a metal-poor ([Fe/H] = $-0.40 \pm
0.04$), solar-type star eclipsed by a low-mass  M-type dwarf in a binary with
an orbital period $P\approx 14.3$\,days and an eccentric orbit
($e\approx0.3$). This eclipsing binary was discovered using photometry from
the WASP project because it shows a transit in the light curve with a depth of
about 2.5\%. \citet{2014A&A...572A..50G} (GMC+2014, hereafter) have presented
a thorough analysis of this binary system based on additional photometry at
optical wavelengths through the primary eclipse (transit) from three
instruments (NITES, OED and BYU) and J-band photometry of the secondary
eclipse obtained with the FLAMINGOS instrument and Kitt Peak National
Observatory  (KPNO). Full details of these observations are given in GMC+2014.

 I have used \texttt{ellc} to analyse these observations of J0113+31. To
explore the model parameter space I used \texttt{emcee}
\citep{2013PASP..125..306F}, a python implementation of an affine invariant
Markov chain Monte Carlo (MCMC) ensemble sampler. The free parameters in the
fit were $r_1$ and $r_2$ -- the radii of the stars relative to the semi-major
axis of the binary, $i$ -- the inclination, $f_s = \sqrt{e}\sin(\omega)$ and
$f_c = \sqrt{e}\cos(\omega)$, $S_{\rm J}$ -- the surface brightess ratio in the
J-band, and $K_1$ -- the semi-amplitude of the primary star's spectroscopic
orbit. The eccentricity and orientation of the orbit are described by the
parameters $f_s $ and $f_c $ because these have a uniform prior probability
distribution. Uniform priors were also adopted for the other parmeters in the
analysis. There are 7 free parameters and I used 50 walkers with 1000 chain
steps to calculate their posterior probability distribution. To speed-up the
calculation I only included WASP data within 0.02 phase units of the transit
in the analysis. The NITES light curve was observed at high cadence so I used
\texttt{ellc} to calculate this light curve for 1-in-10 of the observed data
points and then used an option implemented in \texttt{ellc} to interpolate the
light curve to other times of observation. The walkers were initialised using
randomly selected parameter values from Gaussian distributions with mean and
standard deviation set from the results of some test runs of \texttt{emcee}.
The convergence of the chain was judged ``by-eye'' by inspection of the
parameter values and the log likelihood as a function of step number. The
best-fit values and standard errors of selected model parameters and derived
values given in Table~\ref{FitTable} were calculated from the median and
standard deviation of parameter values in the chain excluding the first 500
steps of each walker. The joint posterior probability distributions for the
parameters are shown in Fig.~\ref{J0113+31_corner}. Further details of the
analysis can be found by inspection of the python script used to generate
these figure that is included with the software distribution.

 The total number of chain steps is 350\,000 and the acceptance fraction for
each walker is $\approx 0.5$ so the number of simulated data sets calculated
to produce these results is  $\approx 700\,000$. The calculation took  17
minutes using 4 threads on a MacBook Pro with a 2~GHz Intel Core i7 CPU. For
comparison, GMC+2014 used a Linux cluster with over 6,000 processor cores and
a peak performance of roughly 24.5 TFLOPS to sample 36\,972  combinations of
four parameters ($i$, effective temperature ratio $T_1/T_2$, $r_1$ and $r_2$)
using the Wilson-Devinney binary star model within the PHOEBE software package
\citep{2005ApJ...628..426P, 1979ApJ...234.1054W}. They did not include the
WASP and NITES photometry in their analysis. The computation of 10\,000 light
curves with this cluster took about 6~hours (Hebb, priv. comm.)

 The agreement between my results and those of GMC+2014 is excellent  and the
precision of radius measurements has been much improved. The small change in
the value of $K_1$ may be the result of an offset between radial velocity
measurements obtained with two different instruments that has not be included
in my analysis. It is beyond the scope of this paper to interpret these
results (which are consistent with those from GMC+2014) and further analysis
is required to determine the sensitivity of these results to important details
such as the adopted limb-darkening coefficients. Nevertheless, it is clear
that \texttt{ellc} makes it feasible to employ Monte Carlo methods to analyse
light curves and radial velocity data for eclipsing binary stars and that it
runs much faster than other binary star models when applied to binary stars
with eccentric orbits. 

\begin{table}
 \caption{Parameters for the eclipsing binary system J0113+31 using
\texttt{ellc} and \texttt{emcee}. Results in the column headed GMC+2014 are
from \citet{2014A&A...572A..50G}. The secondary eclipse depth 
is $\Delta F_{\rm sec} = S_J\left(r_2/r_1\right)^2$.
\label{FitTable}}
\begin{center}
 \begin{tabular}{lrrl}
\hline
  \multicolumn{1}{l}{Parameter} &
  \multicolumn{1}{l}{This work} &
  \multicolumn{1}{l}{GMC+2014 } \\
\hline
\hline
 \noalign{\smallskip}
$r_1$               & $ 0.0533 \pm 0.0004   $ & $ 0.0534 \pm 0.0021  $ \\
$r_2 $              & $ 0.00783 \pm 0.00008 $ & $ 0.0081  \pm 0.0004 $ \\
$i~[^{\circ}] $     & $ 89.09 \pm 0.05      $ & $ 89.08 \pm 0.04     $ \\
$e$                 & $ 0.3096  \pm 0.0007  $ & $ 0.3098 \pm 0.0005  $ \\
$\omega [^{\circ}]$ & $ 278.9 \pm 0.03      $ & $ 278.85 \pm 1.29    $ \\
$\Delta F_{\rm sec}$& $ 0.00749\pm 0.00018  $ & $ 0.00737\pm 0.00024 $ \\
$K_1$ [km/s]        & $ 15.87 \pm 0.02      $ & $ 15.84 \pm 0.01     $ \\
 \noalign{\smallskip}
\hline
 \end{tabular}   
\end{center}
 \end{table}

\begin{figure}
\resizebox{\hsize}{!}{\includegraphics{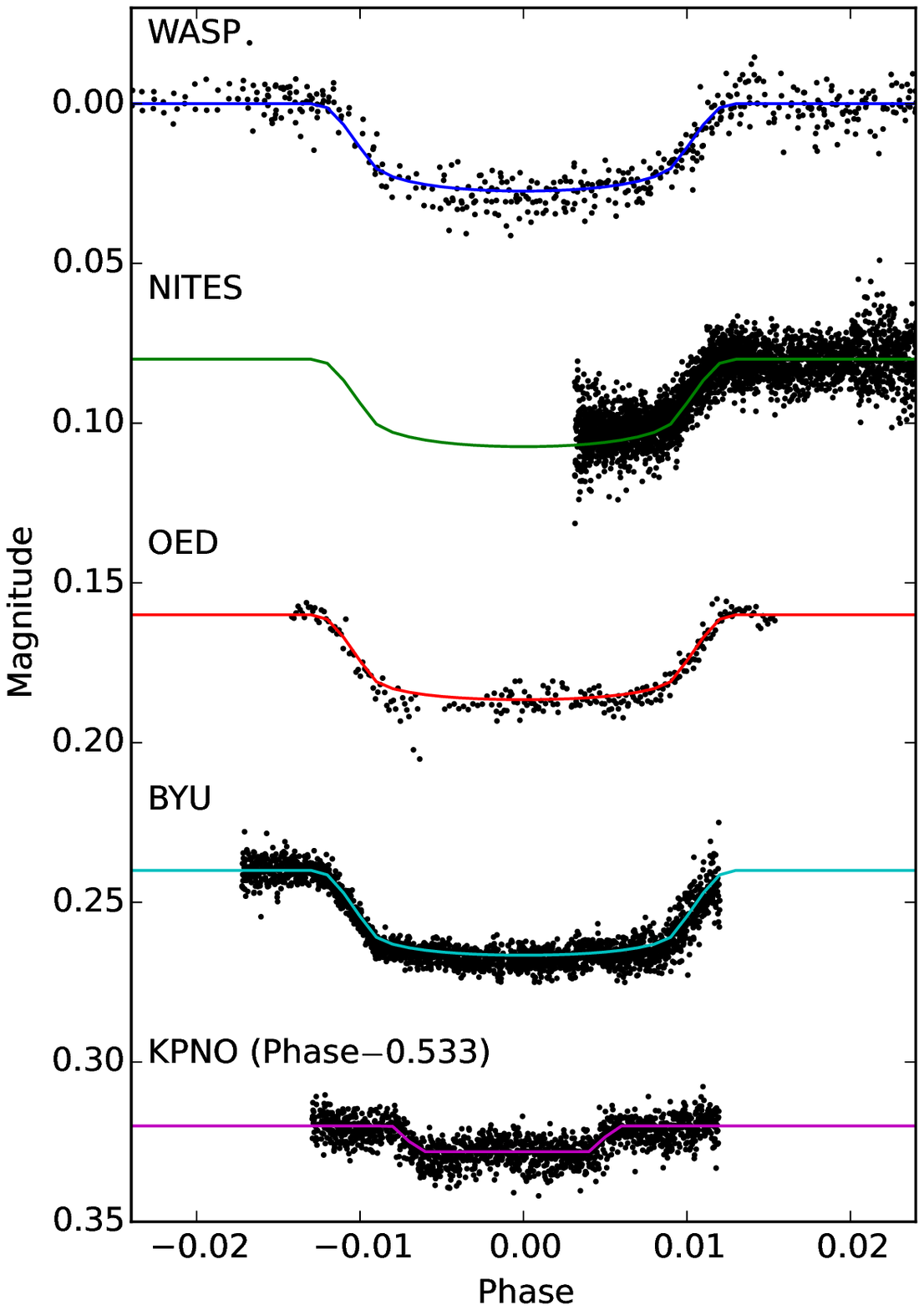}}
\caption{Observed light curves for the eclipsing binary J0113+31
\citep[points, ][]{2014A&A...572A..50G}
compared to  light curves generated using \texttt{ellc} for the parameters
shown in Table~\ref{FitTable} (solid lines).
\label{J0113+31}}
\end{figure}

\begin{figure*}
\resizebox{\hsize}{!}{\includegraphics{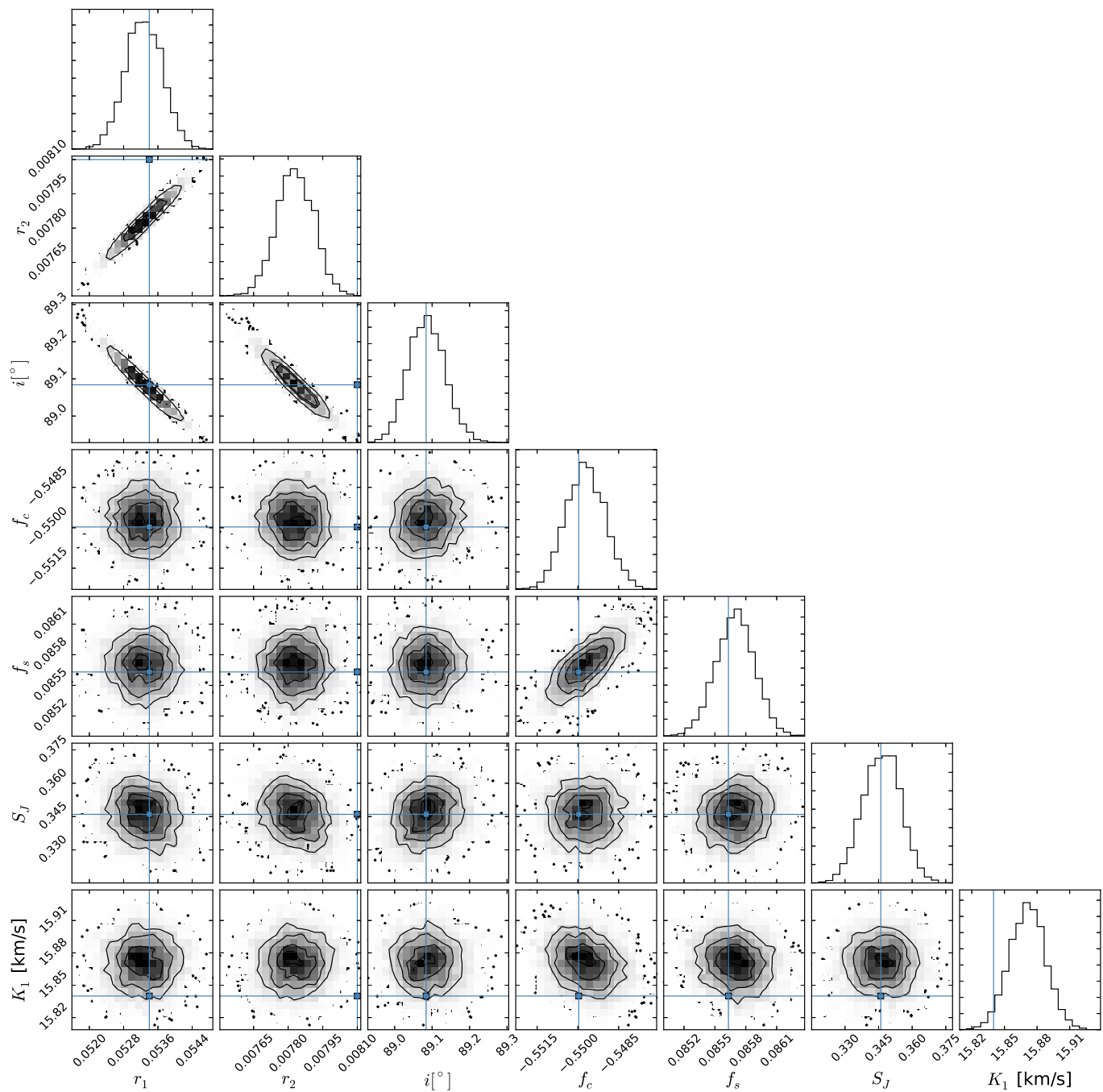}}
\caption{Parameter correlation plots for my fit using \texttt{ellc} to
the radial velocity data and light curves of the eclipsing binary J0113+31
from \citet{2014A&A...572A..50G} (GMC+2014). The adopted parameter values from
GMC+2014 are indicated by lines/points in each panel. The contours in each
panel show the 0.5-, 1-, 1.5- and 2-sigma confidence regions for each joint
parameter distributions. Points outside the 2-sigma confidence region are
individual steps from the \texttt{emcee} MCMC chain. 
\label{J0113+31_corner}}
\end{figure*}

\subsection{HD 23642}

 The \texttt{ellc} package includes a python script called
\texttt{ellc\_emcee.py} that can be used to
analyse a single light curve using the \texttt{ellc} binary star model. The
best-fit light curve and the posterior probability distribution of the free
parameters in the model are calculated using the  \texttt{emcee} algorithm.
The script prompts the user for the values of the fixed and free parameters
and any priors to be imposed on the parameters of the problem. Output file
names and other options such as the grid size can be set using command-line
options. A log file is produced that includes a copy of the user input that
can modified and then  used directly as input into subsequent runs of
\texttt{ellc\_emcee.py}, e.g., for reanalysis of the same light curve with a
different set of free parameters or different starting values for free
parameters. Up to 2 spots per star can be included in the model.
Limb-darkening and gravity-darkening are specified using the effective
temperature and surface gravity of each star and the metallicity of the binary
system. These parameters are then used to look-up the  limb-darkening
coefficients for a 4-parameter limb darkening law and the gravity-darkening
coefficients for each star using linear interpolation in the tabulation of
\citet{2011A+A...529A..75C}. Options to exclude data from the fit, interpolate
the model across data, or to use numerical interpolation to account for the
exposure time are all set using an integer flag value in the input light curve
data. 

 To demonstrate some of the capabilities of  \texttt{ellc\_emcee.py} I have
used this script to analyse the Kepler K2 light curve of HD\,23642. This is an
eclipsing binary in the Pleiades star cluster with an orbital period of about
2.5 days showing partial eclipses between two A-type stars
\citep{2005A+A...429..645S, 2016arXiv160201901D}.  I used the Kepler K2
data corrected for systematic errors using the algorithm of
\citet{2014PASP..126..948V} for this
analysis.\footnote{\url{https://archive.stsci.edu/k2/}}  To speed-up the
calculation I used interpolation to calculate the model values for
approximately 3/4 of the observations between the eclipses  and used the
``very\_sparse'' grid size ($N_j$ = 4) for the numerical integration of the
fluxes from both stars. For observations through the eclipses numerical
integration using 5 integration points was used to account for the integration
time of 1766\,s. The residuals from a preliminary fit to the data show that
the scatter around the best fit is quite variable with time -- there is a
block of very good data with slightly noisier data either side -- so I set the
estimated standard error of each observation to one of three values according
to the time of observation. The values were choosen to achieve a reduced
$\chi^2$ value for the best fit model close to 1. I excluded 10 discrepant
observations from the analysis. The MCMC chain was calculated using 40 walkers
and 5000 steps. This calculation took {\bf 55 minutes} using 8 threads on
a MacBook Pro with a 2~GHz Intel Core i7 CPU.

\begin{figure*}
\resizebox{\hsize}{!}{\includegraphics{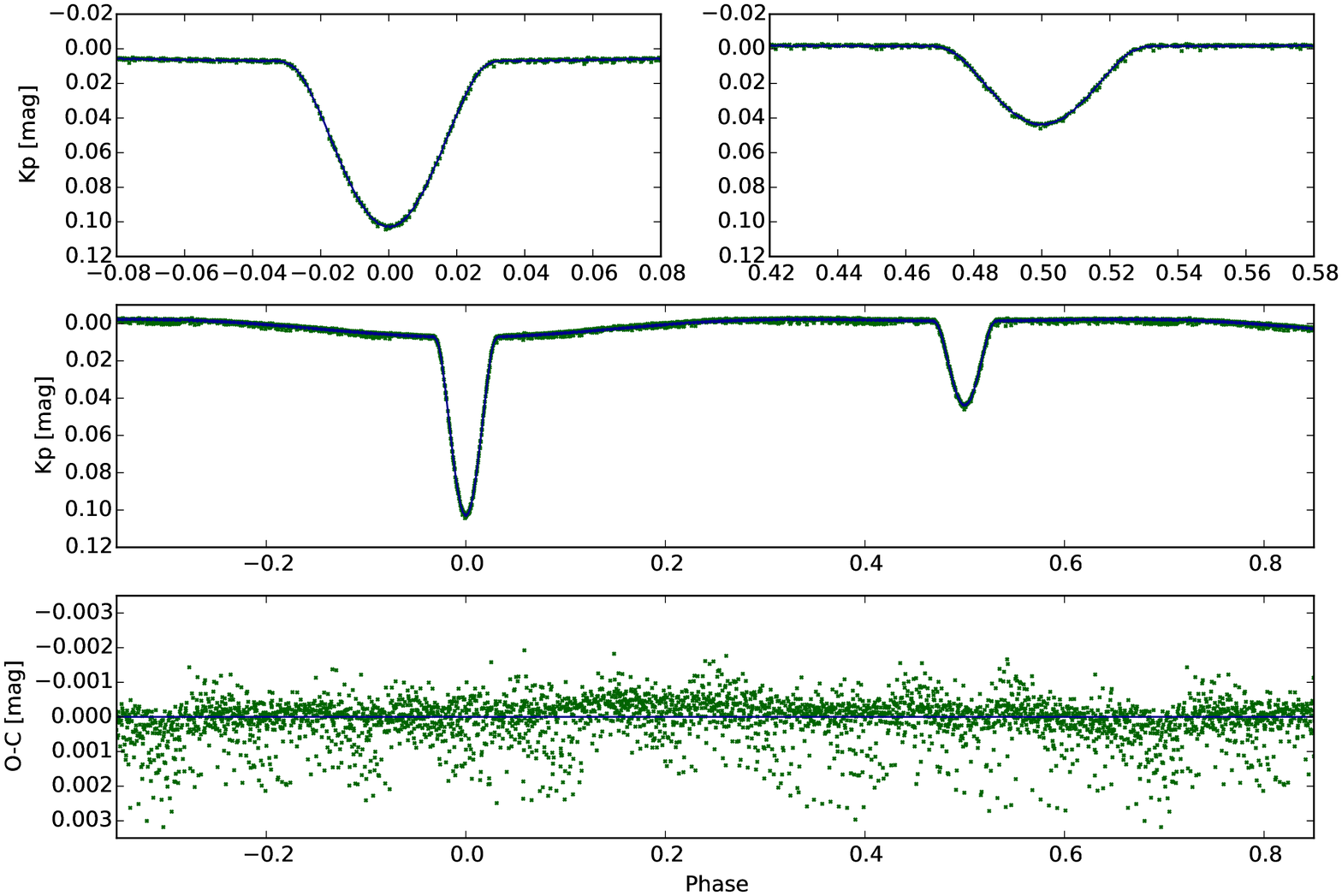}}
\caption{Observed light curves for the eclipsing binary HD\,23642
(points) compared to  light curves generated using \texttt{ellc\_emcee.py} for
the parameters shown in Table~\ref{HD_FitTable} (solid lines).
\label{HD_lcfit}}
\end{figure*}

\begin{figure*}
\resizebox{\hsize}{!}{\includegraphics{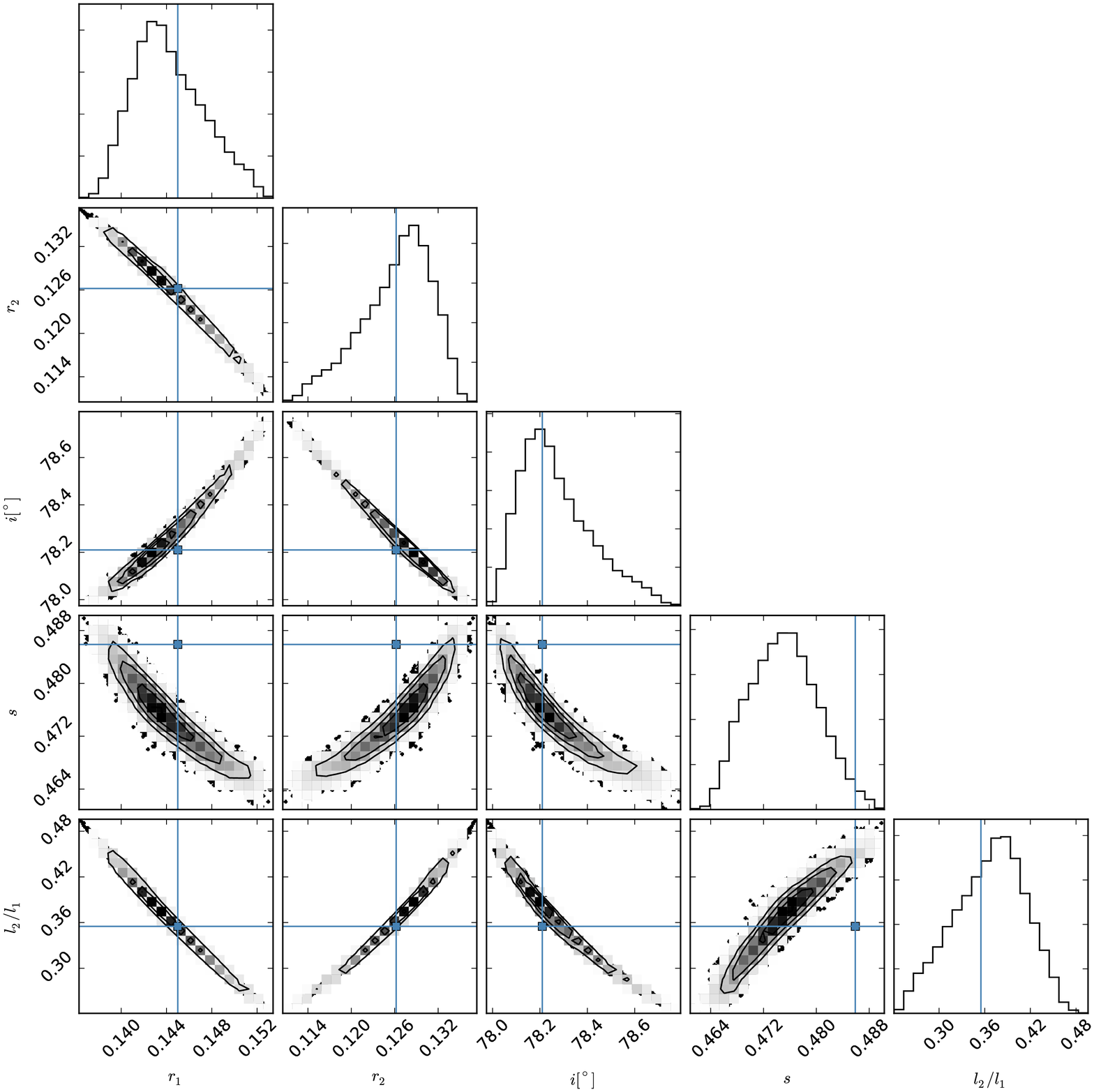}}
\caption{Parameter correlation plots for selected parameters from my fit
using \texttt{ellc\_emcee} to the Kepler K2 light curve of HD~23642. The
parameter values from the analysis of \citet{2016arXiv160201901D} are indicated
by lines/points in each panel. The contours in each panel show the 0.5-, 1-,
1.5- and 2-sigma confidence regions for each joint parameter distributions.
Points outside the 2-sigma confidence region are individual steps from the
\texttt{emcee} MCMC
chain. \label{HD_corner}}
\end{figure*}

 The best-fit parameters  and their standard errors are shown in
Table~\ref{HD_FitTable}. These were calculated using the median and standard
deviation of approximately  145\,000 values in the MCMC chain after
discarding the first $N_{\rm burn-in}$ points from each walker, where  $N_{\rm
burn-in}$ is $4\times$ the largest autocorrelation length scale of the chains
for all the parameters for each walker. The best-fit light curve model is
shown in Fig.~\ref{HD_lcfit} and the distributions of selected parameters are
shown in Fig.~\ref{HD_corner}. Further details of the analysis can be found by
inspection of the files included in the sub-directory
\texttt{examples/ellc\_emcee}  that is included with the software
distribution.

 The agreement between my results and the analysis by
\citet{2016arXiv160201901D} using \texttt{jktebop} \citep{2013A&A...557A.119S}
is very good. There is a  slight difference between the surface brightness
ratio derived here and by \citet{2016arXiv160201901D}, partly because
\texttt{ellc} defines this ratio for the stars when viewed at conjunction,
whereas the \texttt{ebop} model used by \citeauthor{2016arXiv160201901D}
defines this quantity at quadrature.

\begin{table*}
 \caption{Parameters for the eclipsing binary system HD\,23642 using
\texttt{ellc\_emcee}. T$_{\rm ld,1}$ and T$_{\rm ld,2}$ are the parameters
used to interpolate the limb-darkening coefficients for stars 1 and 2 from the
tabulation of \citet{2011A+A...529A..75C}. The time of primary eclipse, $t_0$
is given as BJD-2457000.
\label{HD_FitTable}}
\begin{center}
 \begin{tabular}{lrrl}
\hline
  \multicolumn{1}{l}{Parameter} &
  \multicolumn{1}{l}{This work} & 
  \multicolumn{1}{l}{\citet{2016arXiv160201901D}}&
  Notes \\
\hline
\hline
 \noalign{\smallskip}
$t_0$             & 119.52218 $\pm$   0.00006 & 119.52207 $\pm$   0.00002 
 & BJD-2457000 for mid-primary eclipse.\\
P [days]          & 2.4611361 $\pm$ 0.0000048 & 2.4611341 $\pm$ 0.0000005 \\
$r_1=R_1/a$       &    0.1435 $\pm$    0.0032 &    0.1450 $\pm$    0.0023 \\
$r_2=R_1/a$       &    0.1274 $\pm$    0.0050 &    0.1262 $\pm$    0.0037 \\
$S_{{\rm K}_p}$   &    0.4756 $\pm$    0.0053 &    0.4859 $\pm$    0.0068 
& Surface brightness ratio \\ 
$i~[^{\circ}] $   &     78.28 $\pm$      0.15 &     78.21 $\pm$      0.11 \\
$\ell_2/\ell_1$   &     0.38  $\pm$     0.05  &     0.355 $\pm$     0.035
& Both studies impose prior $\ell_2/\ell_1=0.354 \pm 0.035$ \\
$\alpha_1$        &     =0.35                 &     0.117 $\pm$     0.002 &
Reflection coefficient\\
$\alpha_2$        &     0.344 $\pm$     0.008 &     0.407 $\pm$     0.007&
Reflection coefficient \\
T$_{\rm ld,1}$ [K]&     11600 $\pm$       900 &                        -- &
Limb-darkening look-up parameter\\
T$_{\rm ld,2}$ [K]&      8150 $\pm$       160 &                        -- &
Limb-darkening look-up parameter\\
$\frac{dm}{dt}\:[10^{-6}]$&16.4 $\pm$     0.7 &                        --
& Linear trend in magnitude \\
$\sigma$ [mmag]   &                      0.67 &                      0.67 &
 Root mean square residual\\
 \noalign{\smallskip}
\hline
 \end{tabular}   
\end{center}
\end{table*}     

\section{Caveats}
\subsection{Radius and volume}
 The values of $A$, $B$ and $C$ that are used to specify the shape and size of
the triaxial ellipsoid used in the calculation of the light curves are set by
requiring that the volume of the ellipsoid is equal to the volume of a sphere
with a radius $R$ specified by the user, i.e. $ABC = R^3$. For
highly-distorted configurations this means that the volume of the star will
not be the same as the volume of the  triaxial ellipsoid used to approximate
its shape. The correct way to deal with this difference will depend on the
context in which \texttt{ellc} is being used. For example, if \texttt{ellc} is
being used to perform a least-squares fit to a total eclipse of a star by a
highly-distorted companion, it may be useful to calculate a correction based
on the ratio of the projected area of a triaxial ellipsoid and an
equipotential in the Roche potential viewed at mid-eclipse. Comparing the
results obtained using \texttt{ellc} for spherical stars to those obtained for
a Roche potential or a polytrope (as was done in Fig.~\ref{batman}) will give
an upper limit to the size of any correction of this sort and so enable the
user to determine if any correction is necessary. 

\subsection{Radial velocities}
 The flux-weighted radial velocity calculated using \texttt{ellc} may not
accurately reproduce high-precision radial velocity measurements because some
effects are not included and the definition of the Doppler shift is ambiguous
if the line profiles that are measured are not symmetric. Effects missing from
\texttt{ellc} include the tranverse Doppler effect and gravitational redshift.
Asymmetric line profiles can be caused by convective blue-shift in cool stars,
the R-M effect, pulsations, star spots, and instrumental effects. If the line
profile is asymmetric then different methods to determine the Doppler shift
will give different results, e.g., a best-fit Gaussian profile will give a
different result to a value based on the maximum of a cross-correlation
function. The size of these systematic errors  will be comparable to the
line-width and can only be properly accounted for by modelling both the
observed line profile and the method used to measure the radial velocity.

\subsection{Star spots}
 The light curves for stars with spots are calculated for a fixed value of the
inclination ($i_0$) and are then corrected for the effects of the eclipses.
This means that light curves for spotted stars in binary systems with varying
inclination will not be reliable. There are several approximations used to
enable the efficient calculation of light curves and radial velocities for
stars with  spots in \texttt{ellc}. Detailed modelling of spotted stars is a
problem well beyond the scope of what can be done with \texttt{ellc} or any
other binary model that uses ``dark circles'' to approximate the complex
phenomena associated with magnetic activity on cool stars. Nevertheless, the
star spot model in \texttt{ellc} can be used to quanitify approximately the
likely effects of star spots on the light curves of  cool stars if the user is
aware of the limitations of the model.

\subsection{Irradiation} No account is made
for the irradiation of a star due to the its own radiation being re-emitted by
its companion. Extreme values of the parameter $H_1$ that determines the
pattern of emission due to irradiation  produce numerical noise in the light
curves because numerical integration schemes do not produce accurate results
if the integrand varies rapidly on a scale smaller than the grid point
separation.

\subsection{Bugs}
 The examples above show that \texttt{ellc} can be used to calculate accurate
light curves and radial velocity curves for a wide range of binary systems.
The tests described here have also highlighted some problems with the current
versions of some existing binary star models. Similarly, there will certainly
be combinations of input parameters for which \texttt{ellc} will not produce
reliable results. Users of \texttt{ellc} are strongly advised to test the
results produced by \texttt{ellc} against other binary star models to check
that they are reliable. The \texttt{ellc} binary star model will be made
available as an open-source software project so that users can contribute to
its development, submit bug reports and access the latest version of the
software.\footnote{\url{https://pypi.python.org/pypi/ellc/}}

\section*{Acknowledgements} 
 The coefficients of the quartic equation for the intersections of two
ellipses are taken from  a document ``The Area of Intersecting Ellipses'' by
David Eberly that was published on the web site www.geometrictools.com,
November 20, 2010. I am grateful to the authors of the software packages
\texttt{emcee}, Nightfall and \texttt{batman} for their generosity in making
their software available. The author gratefully acknowledges everybody who has
contributed to make the Kepler Mission possible. Funding for the Kepler
Mission is provided by NASA’s Science Mission Directorate. I thank the referee
for their comments on the manuscript that have helped to clarify a few points
in the text and that also led to a useful update to the software. This work
was supported by the Science and Technology Facilities Council
[ST/M001040/1].

\bibliographystyle{aa} 
\bibliography{wasp}

\begin{appendix}
\section{The intersections of two ellipses}
 A pair of ellipses can be specified using  quadratic equations of the form
\[s_0 +s_1x+s_2y+s_3x^2 +s_4xy+s_5y^2 =0\]
and
\[t_0 +t_1x+t_2y+t_3x^2 +t_4xy+t_5y^2 =0.\]
The intersections of the ellipses are then simultaneous solutions of both
quadratic equations, i.e. solutions of a quartic equation of the form
\[u_0 +u_1y+u_2y^2 +u_3y^3 +u_4y^4 = 0,\]
and similarly for $x$. The coefficients of the quartic equation in $y$ are
\[u_0 = d_{31}d_{10}-d^2_{30}, \] 
\[u_1 = d_{34}d_{10} + d_{31}(d_{40} + d_{12}) - 2d_{32}d_{30}, \]
\[u_2 = d_{34}(d_{40} + d_{12}) + d_{31}(d_{42} - d_{51}) - d^2_{32} - 
2d_{35}d_{30},  \]
\[u_3 = d_{34}(d_{42} - d_{51}) + d_{31}d_{45} - 2d_{35}d_{32}, \]
\[u_4 = d_{34}d_{45} - d^2_{35} ,\]
where $d_{ij} = s_i t_j - s_jt_i$.  Equivalent expressions for a quartic
equation in $x$ can be found by exchanging the indices $1 \leftrightarrow 2$
and $3 \leftrightarrow 5$. 

\begin{figure}
\resizebox{\hsize}{!}{\includegraphics{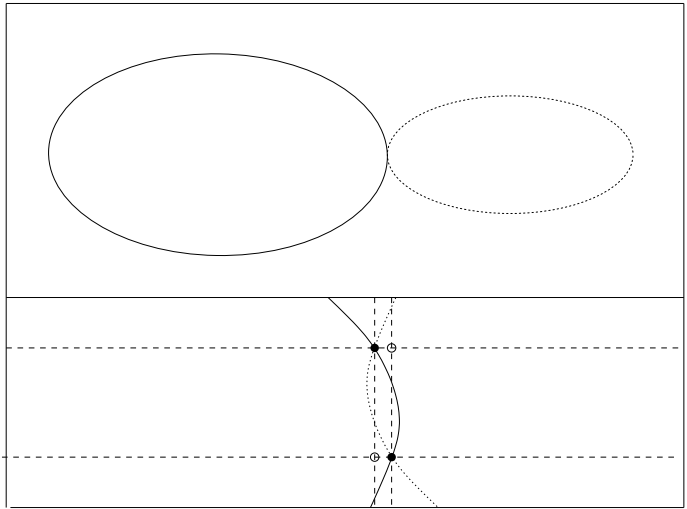}}
\caption{The intersection of two ellipses in a configuration that makes the
calculation of the intersections difficult using floating point arithmetic.
The lower panel shows a sketch of an exaggerated view near the intersection
region. The dashed lines show the values of $x$ and $y$ that should be
obtained by solving the relevant quartic polynomials.  Dots mark the
intersection points and open circles mark spurious solutions of both quartics.
\label{badellipsefig}}
\end{figure}

 Various closed-form solutions of quartic polynomial equations have been
published since the first by \citet{Cardano} but it appears that no algorithm
has been developed that is completely robust when implemented using floating
point arithmetic \citep{Herbison-Evans}. Instead, \texttt{ellc} uses the
\texttt{rpoly} algorithm \citep{Jenkins:1975:AZR} followed by
``root-polishing'' using Laguerre's algorithm \citep{1992nrfa.book.....P} to
ensure that the correct number of real roots is found and to ensure that they
are accurate to within the machine floating-point precision.  Given the roots
of the quartics in $x$ and $y$, the remaining task is to identify which
combinations of these roots correspond to intersections of the ellipses. This
proves to be quite problematic because of cases such as those shown in
Fig.~\ref{badellipsefig}, which often occur in the calculation of the eclipses
for two stars with inclinations close to $90^{\circ}$. These cases also suffer
from inaccurate values of the coefficients $u_0 \dots u_4$ because of rounding
errors. The roots of the quartics will give inaccurate estimates of the
positions of any intersections in these cases no matter how good the
root-finding algorithm. This makes it extremely difficult to find a robust way
to determine if some, none or all of the combinations of roots correspond to
actual intersections, tangent points (which I do not consider to be
intersections in this context), ``near-misses'' or spurious solutions of the
quartics that correspond to incorrect combinations of $x$ and $y$.  

 Most of these problems are the result of the relative orientations of the
ellipses, so to find which combinations of $x$ and $y$ correspond to actual
intersections the coordinate system is rotated by $45^{\circ}$, the quartic
equations for $x^{\prime}$ and $y^{\prime}$ are solved in the transformed
coordinate system, and then apply the inverse transformation to the resulting
real roots. Only combinations of $x$ and $y$ that correspond to intersection
points or tangent points will appear in the lists of paired real roots
calculated for both orientations.  Since it is not known {\it a priori} which
of the two orientations gives reliable results, this analysis is repeated
for a third orientation rotated by $60^{\circ}$ to the first and search for
combinations of $x$ and $y$ that are common to at least two orientations. The
resulting list of candidate intersection points is then searched for
duplicates with a tolerance  of $10^{-9}$, which are then removed.  If only
one candidate intersection point remains then this must be a tangent point. If
two unique candidate intersection points are found a simple test of whether
the mid-points between these points on the circumference of one ellipse lie
inside or outside the other ellipse is sufficient to determine if one ellipse
is tangent at two points to the other. A similar test can be used to identify
the tangent point in the case of 3 candidate intersection points.

\end{appendix}

\end{document}